# Gradient Nonlinear Pancharatnam-Berry Metasurfaces

Mykhailo Tymchenko, J. Sebastiàn Gomez-Diaz, Jongwon Lee, Nishant Nookala, Mikhail A. Belkin, and Andrea Alù[*]

Department of Electrical and Computer Engineering, The University of Texas at Austin, Austin, TX, USA

*alu@mail.utexas.edu

*We apply the Pancharatnam-Berry phase approach to plasmonic metasurfaces loaded by highly nonlinear multi-quantum well substrates, establishing a platform to control the nonlinear wavefront at will based on giant localized nonlinear effects. We apply this approach to design flat nonlinear metasurfaces for efficient second-harmonic radiation, including beam steering, focusing, and polarization manipulation. Our findings open a new direction for nonlinear optics, in which phase matching issues are relaxed, and an unprecedented level of local wavefront control is achieved over thin devices with giant nonlinear responses.*

Artificially engineered metasurfaces have recently attracted a great deal of interest due to their ability to provide a large degree of control over the local amplitude, phase, and polarization of local fields, leading to many exciting advances in science and technology [1,2]. Conventional optical devices are based on the naturally weak interactions of light with matter, implying that volumetric effects dominate their optical response. Metasurfaces provide an elegant way to overcome these constraints, by manipulating the local field with suitably engineered inclusions that can enhance the local interaction with light, and pattern it in the desired way over subwavelength distances. Artificially engineered metasurfaces have for the most part been limited to their linear operation to date, with numerous applications such as wave front engineering [1–4], information processing and analog computations [5], spin-orbit manipulation [6], and three-dimensional holography [7], among many others. Artificially engineered metasurfaces have started to make their way into nonlinear optics, where they hold a great promise to reduce the size and dimensionality of current devices, relax issues associated with phase matching requirements [8,9], and boost the nonlinear response [10,11].

Recently, planar ultrathin nonlinear metasurfaces based on the strong coupling of plasmonic resonances with intersubband transitions of multi-quantum-well (MQW) semiconductor heterostructures have been shown to produce nonlinear responses that are orders of magnitude larger than natural nonlinear crystals with similar thicknesses [10,11]. MQW heterostructures are known to provide one of the largest nonlinear responses in condensed matter, however they respond only to electric fields oriented normally to the semiconductor layers [12–18]. This problem was successfully addressed by employing properly designed plasmonic structures that support highly confined resonances at pump and generated frequencies, efficiently coupling the impinging beam to electric field components perpendicular to the semiconductor layers. The giant level of nonlinearities experimentally observed in these systems open a new paradigm in nonlinear optics, because they can engage very large nonlinear responses in deeply subwavelength volumes, relaxing the necessity for phase matching, and providing significant nonlinear response in a confined pixel. This feature ideally lends itself to the possibility of creating metasurfaces able to control the generated nonlinear fields by gradually varying their local phase and amplitude with subwavelength resolution, allowing the use of reflectarray concepts for wavefront engineering of the nonlinear generated beam [1,19,20].

In the realm of linear metasurfaces, there have been several approaches to control the amplitude and phase of the transmitted wavefront, such as locally changing the size of metallic inclusions, or the apex angle of V-shaped nanoantennas, or placing metallic particles in elaborate periodic and aperiodic arrangements, to name a few [1]. Another approach is to employ optical elements suitably designed based on the Pancharatnam-Berry (PB) phase concept (PB optical elements), which introduce a topological phase difference between transmitted (or reflected) waves based on their geometry and orientation [21,22]. It has been shown that flat metasurfaces consisting of PB optical elements can efficiently tailor the local transmitted or reflected wave by gradually varying their local orientation from cell to cell, thus enabling wide wavefront engineering capabilities over a flat platform [23–25]. In this approach, the inclusions need to be accurately designed to ensure proper coupling with the polarization of interest, and to minimize other diffraction effects.

Here, we extend the concept of PB-phase optical elements to the nonlinear regime, and more specifically to MQW-based nonlinear plasmonic metasurfaces, in order to tailor at will the spatial phase distribution of their efficient second harmonic radiation [10]. Since these giant nonlinear effects are very sensitive to variations in the local resonances of the metasurface, the PB phase approach becomes an ideal tool to achieve full phase control and at the same time ensures nearly uniform, giant nonlinear response across the metasurface, based on a single suitably designed PB element that gradually changes its orientation from cell to cell. In addition, our numerical analysis (see [28]) provides an ideal tool to compute both the *amplitude* and *phase* of the emerging SH fields, irrespectively of the resonator design or the non-linear material employed. This allows the fast design of nonlinear plasmonic metasurfaces with advanced functionalities such as light bending or focusing, while simultaneously providing conversion efficiencies several orders of magnitude larger than any other planar nonlinear configuration [10]. Very recent attempts in this direction [26,27], of which we were unaware at the time of submission, have indeed shown that the concept of gradient metasurfaces can be applied to nonlinear optics. In the following, we apply this concept to multi-quantum-well loaded metasurfaces, enabling large conversion efficiency and full control of the phase and amplitude of the generated nonlinear beams with subwavelength resolution.

The general concept of the proposed nonlinear PB metasurface is illustrated in Fig. 1. A thin MQW substrate with layers grown in the *x-y* plane is sandwiched between an array of suitably designed plasmonic resonators and a metallic ground plane. The incident beam propagates along the *z*-direction, and the metasurface operates in reflection. Each element is designed to ensure giant nonlinear response, similar to [11], and more specifically large second-harmonic conversion efficiency, but at the same time to ensure a subwavelength footprint. In addition, here we rotate each element of the surface to acquire the desired local geometrical phase for circularly polarized incidence.

Assuming that the coupling between neighboring elements is weak, we can describe the optical response of each element using its effective local nonlinear transverse susceptibility tensor $\ddot{\chi}^{(2)\,\text{eff}}$, which relates the induced nonlinear transverse polarization density averaged over the volume of the element at frequency $2\omega$ to the transverse incident field at $\omega$ [10]. In order to apply the PB phase approach, we recast this tensor in $\chi^{(2)}_{zzz}$ circular polarization basis as $\chi^{(2)\,\text{eff}}_{\alpha\beta\gamma}$, where $\alpha$, $\beta$, and $\gamma$ can be *R* or *L*, corresponding to RHCP or LHCP fields, respectively [28]. Once this effective susceptibility tensor is known, the averaged transverse nonlinear surface currents induced on the metasurface can be readily obtained. Note that we neglect the *z*-polarized contribution to these currents, which may become relevant for radiation significantly away from broadside (see [28] for the validation of this assumption). For the sake of simplicity, we assume here that only one circularly polarized wave is incident at a time, so that $\gamma = \beta$ always holds. It can be shown that each PB element, rotated as $\varphi(x, y)$ across the surface, when illuminated by a LHCP incident wave $E^{\omega}_{L(\text{inc})}$ at normal incidence, generates an effective nonlinear transverse surface current that can be split into LHCP and RHCP components denoted as $K^{2\omega}_{L(L)}$ and $K^{2\omega}_{R(L)}$, with spatial variation analytically given by [28]

$$K^{2\omega}_{L(L)}(x, y) = 2\omega\varepsilon_0 h \chi^{(2)\,\text{eff}}_{LLL} [E^{\omega}_{L(\text{inc})}]^2 \exp[i3\varphi(x, y)], \tag{1a}$$

$$K^{2\omega}_{R(L)}(x, y) = 2\omega\varepsilon_0 h \chi^{(2)\,\text{eff}}_{RLL} [E^{\omega}_{L(\text{inc})}]^2 \exp[i\varphi(x, y)], \tag{1b}$$

where $h$ is the height of the MQW layer. Similar expressions for a RHCP impinging wave $E^{\omega}_{R(\text{inc})}$ possess the opposite dependence on the local orientation $\varphi(x, y)$, as detailed in [28]. In Eq. (1) we have taken into account that, for a given metasurface configuration, the radiated waves propagate in the direction opposite to the incident wave, and therefore have a reversed circular polarization basis. Remarkably, and differently from the conventional PB phase approach, no optimization of the coupling efficiency to different polarization mechanisms is necessary in this nonlinear operation, and Eqs. (1a), (1b) always apply. The design of PB elements can therefore be focused on maximally enhancing the nonlinear process of interest to realize giant nonlinear response, and once the optimal inclusion is selected, it simply needs to be rotated gradually along the surface. In addition, for each circularly polarized normally incident wave, the nonlinear output is automatically split into a pure circularly polarized basis, with different patterning depending only on the local orientation of the inclusions on the surface.

In order to demonstrate the aforementioned concepts, we first design an optimized unit cell to enhance local second-harmonic generation, and then we apply this design to realize PB metasurfaces with tailored nonlinear wavefronts. The optimized cell consists of a U-shaped split-ring plasmonic resonator [see Fig. 2(a)], and follows the general requirements determined in [10], i.e., (i) locally enhancing the field at resonance, (ii) supporting resonances with overlapping modal distributions at fundamental and second-harmonic frequency, and (iii) efficiently converting the transverse impinging and outgoing electric fields into locally enhanced vertical fields. In addition, the PB approach imposes additional requirements to the unit-cell design, including that (iv) it must sit on a subwavelength footprint, (v) it must allow rotation within the same footprint, and (vi) it must ensure weak coupling between neighboring cells. In our design, we employ the same MQW heterostructure as in [10], with thickness $h = 500\,\text{nm}$ and $\chi^{(2)}_{zzz} = 54\,\text{nm} \cdot \text{V}^{-1}$ at 37 THz, and we etch the MQW layer around the resonator to reduce the coupling between adjacent PB elements. The size of the plasmonic resonator and unit cell, specified in Fig. 2(b), were optimized to achieve overlapping resonances at fundamental and second harmonic frequencies, 37 THz ($\lambda_\omega = 8\,\mu\text{m}$) and 74 THz ($\lambda_{2\omega} = 4\,\mu\text{m}$), respectively. Fig. 2(c) shows the spatial distribution of the normalized $z$-component of the electric field near the plasmonic resonator at the two frequencies. At the fundamental frequency $\omega$ the structure is efficiently excited by a $y$-polarized wave, whereas at $2\omega$ the structure responds to $x$-polarized fields, as confirmed in the absorption spectrum shown in Fig. 2(d). Our numerical simulations [28] confirms a conversion efficiency above $2 \cdot 10^{-4}\%$, similar to the one experimentally obtained in [10] and several orders of magnitude larger than those found in planar nonlinear metasurfaces based on conventional optical materials.

The first example of nonlinear PB metasurface operation is aimed at steering the generated beams towards specific directions. To this goal, the metasurface should provide a linear phase gradient along one direction at the second harmonic frequency $2\omega$, realizing a periodic superlattice composed of supercells with period $L = Nd$, where $d$ is the size of each unit cell, and $N$ is the number of elements required to complete a full turn around the $z$-axis, i.e., from $\varphi = 0$ to $\varphi = 360$ deg. We choose the unit cell size $d = 1.5$ μm [Fig. 2(b)] with an angular rotation step $\Delta\varphi = 15$ deg between neighboring cells, chosen to be small enough in order to limit unwanted phase variations that may break the assumptions at the basis of Eq. (1). The supercell therefore contains $N = 24$ unit cells, corresponding to $L = 36$ μm. From basic reflectarray theory [19] and Eqs. (1a), (1b), it follows that for a LHCP wave normally incident on the surface at $\omega$, the waves radiated by the LHCP and RHCP currents at $2\omega$ will propagate at angles $\theta_{L(L)} = -\mathrm{asin}[(3\Delta\varphi/360°)\lambda_{2\omega}/d]$ and $\theta_{R(L)} = -\mathrm{asin}[(\Delta\varphi/360°)\lambda_{2\omega}/d]$, which for our geometry are $-20$ deg and $-7$ deg with respect to the $-z$-direction, respectively. Analogously, from Eq. (3a) and (3b) it follows that under RHCP incidence the metasurface supports RHCP and LHCP currents radiating at $\theta_{R(R)} = -\theta_{L(L)}$ and $\theta_{L(R)} = -\theta_{R(L)}$, respectively. It should be noticed that, while the induced currents are purely CP as predicted by Eq. (1), the radiated waves in general are only partially CP, as they travel at an angle from the normal. However, for radiation angles relatively close to the normal, as in the cases considered here, the radiated waves are circularly polarized with very good approximation [28]. Fig. 3(a) shows the analytically and numerically calculated phases of the effective induced surface currents $K^{2\omega}_{\alpha(L)}(x)$ under LHCP illumination, as a function of position along the superlattice period $L$ (see [28] for a description of computational methods). The corresponding variation of the magnitude of the tensor elements $\chi^{(2)\mathrm{eff}}_{\alpha\beta\beta}$ normalized to $\chi^{(2)}_{zzz}$ is shown in Fig. 3(b), reporting average values $\langle \chi^{(2)\,\text{eff}}_{RLL} \rangle = 11 \text{ nm V}^{-1}$, $\langle \chi^{(2)\,\text{eff}}_{LLL} \rangle = 19 \text{ nm V}^{-1}$, which are of the same order as the results reported in [10], confirming that the phase control functionality does not affect the overall high efficiency of the non-linear process (see [28] for an extended discussion on conversion efficiency). The slight discrepancy between theoretical and numerical results are due to the small differences in the coupling between adjacent resonators as a function of their orientation. Fig. 3(c) show the simulated spatial distribution of $E_y^{2\omega}$ above the metasurface illuminated by a 30μm-wide LHCP Gaussian beam. The simulation results confirm that the radiated field cleanly splits into two separate beams with opposite handedness and different directions, as predicted by our theory. In our design, since $\chi^{(2)\,\text{eff}}_{LLL} > \chi^{(2)\,\text{eff}}_{RLL}$, the LHCP beam has larger amplitude than the RHCP one, which is clearly observed in Fig. 3(d). Proper optimization of the unit cell may provide similar intensities, or completely suppress one of the two beams, depending on the application of interest. Fig. 3(d-f) show the results for the same PB metasurface under RHCP excitation. As mentioned above, in this case the phase shifts and directions of the beams are opposite. The average values of the nonlinear susceptibility tensor elements are the same, namely $\langle \chi^{(2)\,\text{eff}}_{RLL} \rangle = \langle \chi^{(2)\,\text{eff}}_{LRR} \rangle$ and $\langle \chi^{(2)\,\text{eff}}_{LLL} \rangle = \langle \chi^{(2)\,\text{eff}}_{RRR} \rangle$. If we excite at oblique incidence, the transverse momentum imprinted on the PB currents is added to the momentum of the impinging excitation, allowing continuous steering of the nonlinear beams. For steeper incidence and radiation angles, additional contributions from vertically polarized nonlinear currents may be expected, depending on the metasurface design, which has been neglected here. We verify and further discuss in [28] how this approximation holds very well for the examples considered in this paper.

Another classical example of linear metasurface operation is focusing the radiated field in the near-field of the metasurface. Since the right- and left-handed polarized components of the second harmonic field possess different phase dependence on the local PB element orientation, their focusing requirements would be different. Our design is aimed at focusing the LHCP component of the generated beam at 20 μm above the nonlinear metasurface under LHCP normal incidence. The required spatial variation of the PB elements orientation is not periodic any longer, but it has a quadratic dependence $\varphi(x) = 360°[(x^2 + f^2)^{1/2} - f]/3\lambda_{2\omega}$ [1]. Fig. 4(a) shows the analytical and numerical phase of the induced surface currents $K_{\alpha(L)}^{2\omega}(x)$. The magnitude of the corresponding $\ddot{\chi}^{(2)\,\text{eff}}$ elements varies from cell to cell slightly more than in the previous case, as can be seen in Fig. 4(b), averaging around $\langle \chi_{LLL}^{(2)\,\text{eff}} \rangle = 24$ nm V$^{-1}$, $\langle \chi_{RLL}^{(2)\,\text{eff}} \rangle = 22$ nm V$^{-1}$. Fig. 4(c) shows the spatial distribution of $E_y^{2\omega}$ for the same 30μm LHCP impinging beam incident at normal incidence, as in the first example. The inset shows the corresponding spatial distribution of the time-averaged energy density of the radiated field. Our results confirm that nearly perfect focusing of the radiated LHCP wave is achieved at the desired point. A change of incident polarization to RHCP will result in strong nonlinear radiation defocusing [28].

In conclusion, we have shown that the PB phase approach constitutes a powerful tool to engineer gradient metasurfaces with giant nonlinear response, greatly enriching their functionality and opening fascinating prospects for wavefront engineering of nonlinear frequency generation, supported by the absence of phase matching requirements. While this approach cannot be considered a direct extension of the linear PB approach, since it cannot be any longer mapped over a single Poincaré sphere, due to the frequency transformation, it also allows to linearly control the amount of phase imparted to the generated nonlinear beam with geometrical rotations. Importantly, and different again from the linear PB approach, the wavefront engineering capabilities of such metasurfaces do not require sacrificing their performance, since the nonlinear response is inherently associated to their subwavelength footprint. Our approach can be easily extended to other nonlinear phenomena, such as third harmonic, sum and difference frequency generation, phase conjugation, and more. In addition, the proposed concept can also be applied to nonlinear metasurfaces with a dielectric substrate that operate in transmission. This work was supported by the AFOSR grant No. FA9550-14-1-0105 and the ONR MURI grant No. N00014-10-1-0942.

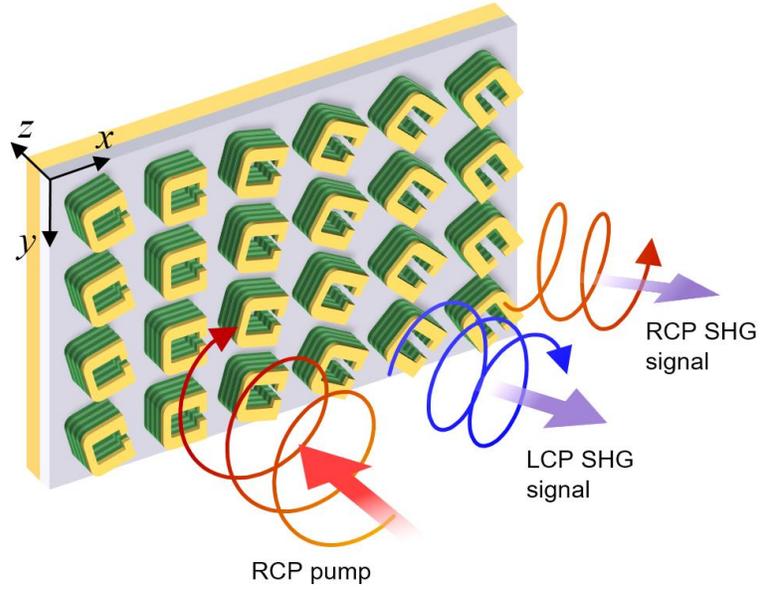

FIG. 1. A sketch of the proposed PB nonlinear metasurface with a phase gradient in the *x*-direction. The MQW blocks are sandwiched between U-shaped gold resonators and a metallic ground plane. The incident circular polarized wave at $\omega$ generates simultaneously RHCP and LHCP nonlinear waves at $2\omega$.

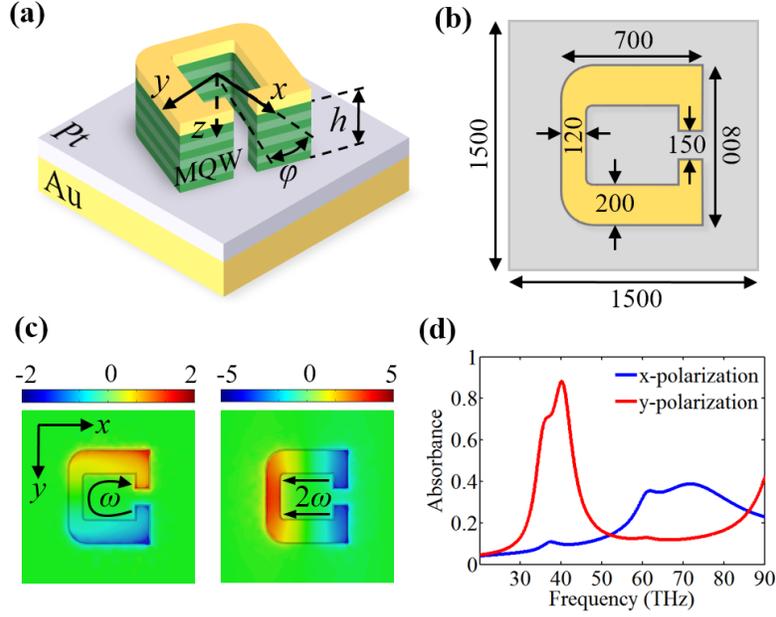

FIG. 2. (a) Geometry of a PB metasurface element. (b) Dimensions (in nm) of the gold plasmonic resonator. The MQW layer is etched around the resonator. (c) Spatial distribution of the normalized $z$-component of the field at the top of the MQW layer at the fundamental and second harmonic frequencies, $E_z^{\omega}$ and $E_z^{2\omega}$, respectively. (d) Simulated absorption spectrum for different polarizations.

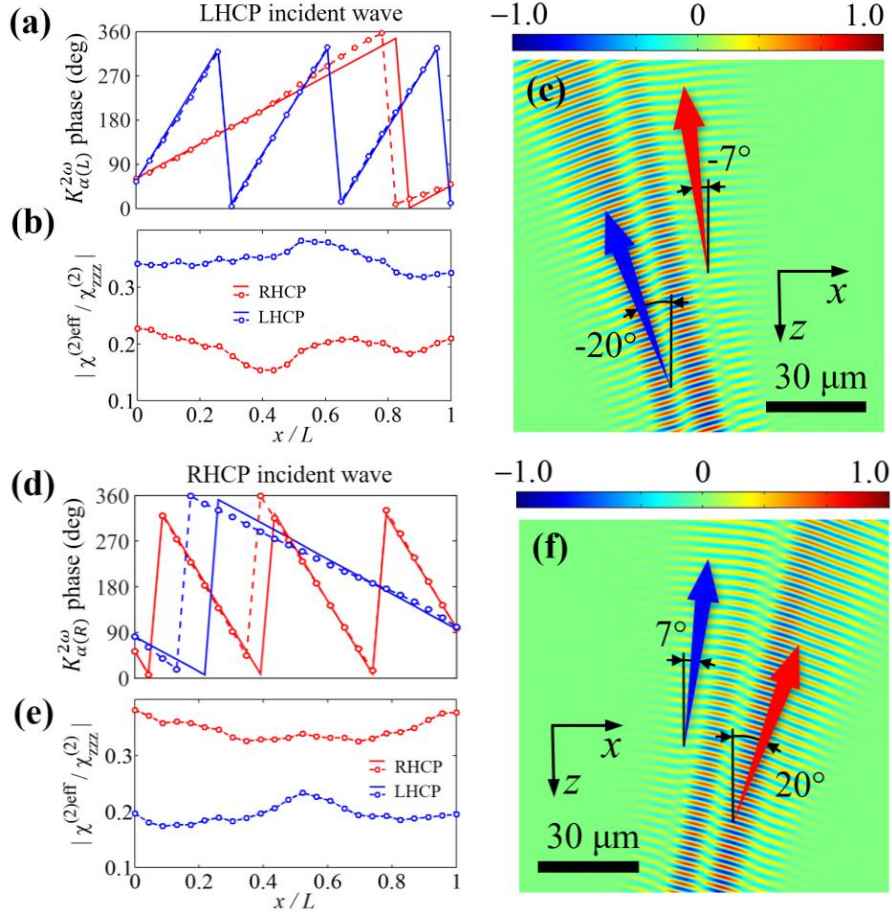

FIG. 3. Analytical and numerical results for a flat nonlinear metasurface with a linear variation of the PB elements' rotation along the *x*-axis, considering an angular step $\Delta\varphi$ of 15 degrees. (a) Phases of RHCP and LHCP components of the effective nonlinear surface current generated on the metasurface by a LHCP plane wave at normal incidence at $\omega$. Analytical results are shown with solid lines, dashed lines with markers show the corresponding numerical results. (b) Magnitude of the $\vec{\vec{\chi}}^{(2)\text{eff}}$ elements computed for each cell, normalized by $\chi_{zzz}^{(2)}$. (c) Spatial distribution of the $E_y^{2\omega}$ component of the radiated field above the metasurface illuminated by a 30μm-wide LHCP Gaussian beam (the incident field is not shown). (d-f) Same as in (a-c), but for an RHCP impinging wave at normal incidence.

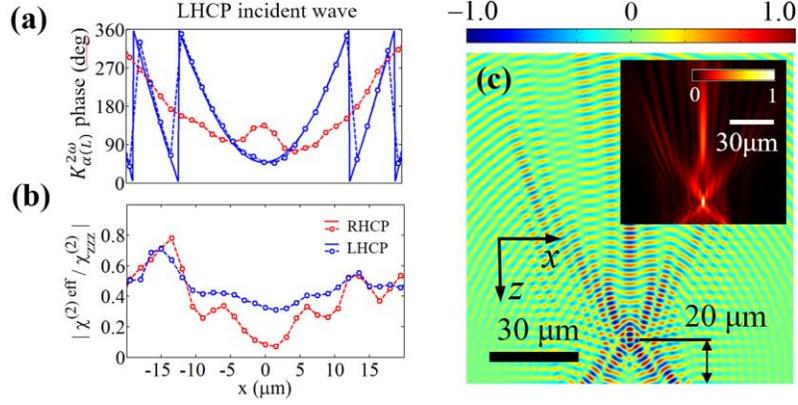

FIG. 4. Focusing the LHCP nonlinear radiation 20 μm above the metasurface. The structure is illuminated by LHCP normally impinging waves. (a) Analytically (solid lines) and numerically (dashed lines with markers) computed phases of RHCP and LHCP components of the effective nonlinear current induced on the metasurface by a LHCP incident plane wave at $\omega$. (b) Local magnitude of the $\vec{\vec{\chi}}^{(2)\text{eff}}$ elements normalized by $\chi_{zzz}^{(2)}$. (c) Spatial distribution of the $E_y^{2\omega}$ component of the radiated field above the metasurface illuminated by a 30μm-wide LHCP Gaussian beam (not shown). The inset shows the corresponding spatial distribution of the energy density.

# Supplemental Information

## I. Geometrical Phase Evaluation

Let us consider a 2D array consisting of subwavelength cells possessing a nonlinear response, designed specifically for efficient second harmonic generation (SHG). We assume that the size of the cell is small enough so that the second-order non-linear response of each cell can be treated in terms of an effective transverse nonlinear susceptibility tensor $\ddot{\chi}^{(2)\text{eff}}$, which relates the induced transverse nonlinear polarization density $\mathbf{P}^{2\omega}$ averaged over the volume of each cell and oscillating at the frequency $2\omega$, and the transverse incident field $\mathbf{E}^{\omega}_{(\text{inc})}$ oscillating at $\omega$. Since the nonlinear multi-quantum-well (MQW) heterostructure considered in the paper has only one non-zero element of its nonlinear susceptibility $\ddot{\chi}^{(2)}$ tensor, $\chi^{(2)}_{zzz}$, it can be shown that in Cartesian coordinates the elements of $\ddot{\chi}^{(2)\text{eff}}$ can be obtained by means of Lorentz reciprocity. In the case of a periodic surface composed by the same elements [10,29,30] it is given by

$$\chi^{(2)\text{eff}}_{ijk} = \frac{1}{V}\int_V \chi^{(2)}_{zzz}(\mathbf{r}) f_i^{2\omega}(\mathbf{r}) f_j^{\omega}(\mathbf{r}) f_k^{\omega}(\mathbf{r}) d^3\mathbf{r}, \qquad (S1)$$

where all the indexes can be $x$ or $y$, $\chi^{(2)}_{zzz}(\mathbf{r})$ is the local value of the intrinsic second-order susceptibility tensor in the unit cell, and $f_i^{\omega}(\mathbf{r})$, $f_i^{2\omega}(\mathbf{r})$ are local enhancement factors of electric field $z$-component $E^{\omega}_{z(i)}(\mathbf{r})$, $E^{2\omega}_{z(i)}(\mathbf{r})$ induced in the unit cell by $i$-polarized normally incident waves $E^{\omega}_{i(\text{inc})}$, $E^{2\omega}_{i(\text{inc})}$, respectively,

$$f_i^{\omega}(\mathbf{r}) = \frac{E^{\omega}_{z(i)}(\mathbf{r})}{\left|E^{\omega}_{i(\text{inc})}\right|}, \qquad f_i^{2\omega}(\mathbf{r}) = \frac{E^{2\omega}_{z(i)}(\mathbf{r})}{\left|E^{2\omega}_{i(\text{inc})}\right|}. \qquad (S2)$$

The integration in (S1) is carried out over the volume $V$ of the MQW layer in the unit cell. For simplicity of notation, in what follows we use $E^{\omega}_{i(inc)} \equiv E^{\omega}_i$, and we also omit the apex "eff" for $\ddot{\chi}^{(2)\text{eff}}$. The induced transverse nonlinear polarization density in each cell is defined as $E^{2\omega}_{z(i)}(\mathbf{r})$

$$P^{2\omega}_i = \varepsilon_0 \sum_{j,k} \chi^{(2)}_{ijk} E^{\omega}_j E^{\omega}_k. \qquad (S3)$$

This homogenization model assumes that $\ddot{\chi}^{(2)\text{eff}}$ does not strongly depend on the direction of impinging waves testing the structure $E^{\omega}_{z(i)}(\mathbf{r})$, , and therefore they are assumed to be impinging at normal incidence, neglecting $z$-oriented effective nonlinear currents, which may become relevant for radiation significantly away from broadside, and the corresponding longitudinal components of the $\ddot{\chi}^{(2)\text{eff}}$ tensor. While this assumption is ideally valid only for periodic surfaces with deeply subwavelength unit cells illuminated normally, in Section II we verify numerically that this assumption is a very good approximation also in the examples considered in the paper, which are not transversely homogeneous. A more rigorous description of the nonlinear surface interactions, including vertical polarization effects and generalizing the following theory to oblique incidence, is possible, but goes beyond the interest of the present paper and will be developed by the authors in the near future.

In order to apply the geometrical phase approach to the described nonlinear problem, it is convenient to work with circularly polarized (CP) waves. Assuming again that the incident wave at frequency $\omega$ impinges from the normal, and that only transverse components of the $\chi^{(2)}_{ijk}$ tensor are relevant, we can perform a coordinate transformation between the linear polarization basis $\{\mathbf{e}_x, \mathbf{e}_y\}$ and the CP basis $\{\mathbf{e}_L^{\pm}, \mathbf{e}_R^{\pm}\}$ corresponding to waves propagating in $\pm z$ direction:

$$\mathbf{e}_L^{\pm} = \frac{1}{\sqrt{2}}(\mathbf{e}_x \pm i\mathbf{e}_y), \qquad \mathbf{e}_R^{\pm} = \frac{1}{\sqrt{2}}(\mathbf{e}_x \mp i\mathbf{e}_y). \tag{S4}$$

The corresponding transformation matrices are

$$\Lambda^{\pm}_{\alpha i} = \frac{1}{\sqrt{2}}\begin{bmatrix} 1 & \pm i \\ 1 & \mp i \end{bmatrix}, \qquad [\Lambda^{\pm}]^{-1}_{i\alpha} = \frac{1}{\sqrt{2}}\begin{bmatrix} 1 & 1 \\ \mp i & \pm i \end{bmatrix}. \tag{S5}$$

The linearly polarized incident wave can be then represented as a superposition of CP waves propagating in $+z$ direction

$$E_i^{\omega} = \sum_{\alpha}[\Lambda^{+}]^{-1}_{i\alpha}E_{\alpha}^{\omega}, \tag{S6}$$

We transform $P_i^{2\omega}$ to the CP basis for radiated waves $\{\mathbf{e}_L^{-}, \mathbf{e}_R^{-}\}$ propagating in $-z$ direction, obtaining

$$P_{\alpha}^{2\omega} = \sum_{i}\Lambda^{-}_{\alpha i}P_i^{2\omega} = \varepsilon_0 \sum_{i,j,k}\Lambda^{-}_{\alpha i}\chi^{(2)}_{ijk}E_j^{\omega}E_k^{\omega}, \tag{S7}$$

where $\alpha, \beta, \gamma = \{L, R\}$. Substituting (S6) into (S7) yields

$$P_{\alpha}^{2\omega} = \sum_{i}\Lambda^{-}_{\alpha i}P_i^{2\omega} = \varepsilon_0 \sum_{\beta,\gamma}\sum_{i,j,k}\Lambda^{-}_{\alpha i}\chi^{(2)}_{ijk}[\Lambda^{+}]^{-1}_{j\beta}[\Lambda^{+}]^{-1}_{k\gamma}E_{\beta}^{\omega}E_{\gamma}^{\omega}. \tag{S8}$$

Now we can define the tensor $\chi^{(2)}_{\alpha\beta\gamma}$ in circular polarization basis

$$\chi^{(2)}_{\alpha\beta\gamma} = \sum_{i,j,k}\Lambda^{-}_{\alpha i}\chi^{(2)}_{ijk}[\Lambda^{+}]^{-1}_{j\beta}[\Lambda^{+}]^{-1}_{k\gamma}, \tag{S9}$$

with components

$$\chi^{(2)}_{RRR} = 2^{-3/2}\left[\chi^{(2)}_{xxx} - \chi^{(2)}_{xyy} - 2\chi^{(2)}_{yxy} + i\left(\chi^{(2)}_{yxx} - \chi^{(2)}_{yyy} + 2\chi^{(2)}_{xxy}\right)\right] \tag{S10a}$$

$$\chi^{(2)}_{LLL} = 2^{-3/2}\left[\chi^{(2)}_{xxx} - \chi^{(2)}_{xyy} - 2\chi^{(2)}_{yxy} - i\left(\chi^{(2)}_{yxx} - \chi^{(2)}_{yyy} + 2\chi^{(2)}_{xxy}\right)\right] \quad \text{(S10b)}$$

$$\chi^{(2)}_{RLL} = 2^{-3/2}\left[\chi^{(2)}_{xxx} - \chi^{(2)}_{xyy} + 2\chi^{(2)}_{yxy} + i\left(\chi^{(2)}_{yxx} - \chi^{(2)}_{yyy} - 2\chi^{(2)}_{xxy}\right)\right] \quad \text{(S10c)}$$

$$\chi^{(2)}_{LRR} = 2^{-3/2}\left[\chi^{(2)}_{xxx} - \chi^{(2)}_{xyy} + 2\chi^{(2)}_{yxy} - i\left(\chi^{(2)}_{yxx} - \chi^{(2)}_{yyy} - 2\chi^{(2)}_{xxy}\right)\right] \quad \text{(S10d)}$$

$$\chi^{(2)}_{RLR} = 2^{-3/2}\left[\chi^{(2)}_{xxx} + \chi^{(2)}_{xyy} + i\left(\chi^{(2)}_{yxx} + \chi^{(2)}_{yyy}\right)\right] \quad \text{(S10e)}$$

$$\chi^{(2)}_{LLR} = 2^{-3/2}\left[\chi^{(2)}_{xxx} + \chi^{(2)}_{xyy} - i\left(\chi^{(2)}_{yxx} + \chi^{(2)}_{yyy}\right)\right] \quad \text{(S10f)}$$

We assume that the cells experience only weak coupling with each other, implying that the rotation of one element does not sensibly affect the resonance of the neighbors. Incidentally, for the structure considered in the main document, the weak coupling is ensured by etching the MQW around the metallic resonators, constraining the strong induced fields to the volume of the semiconductor. As a result, by properly rotating the elements we change the phase of $P_L^{2\omega}$ and $P_R^{2\omega}$ induced in the MQW. More quantitatively, by rotating the structure around the $z$-axis by an angle $\varphi$, so that the rotation matrix $R_{ij}(\varphi)$ is given as

$$R^+_{ij} = [R^-]^{-1}_{ij} = \begin{bmatrix} \cos\varphi & \sin\varphi \\ -\sin\varphi & \cos\varphi \end{bmatrix}, \qquad [R^+]^{-1}_{ij} = R^-_{ij} = \begin{bmatrix} \cos\varphi & -\sin\varphi \\ \sin\varphi & \cos\varphi \end{bmatrix}, \quad \text{(S11)}$$

in circular polarization basis the rotation matrix acting on the vector $\mathbf{E} = (E_L, E_R)$ becomes

$$R^+_{\alpha\beta} = [R^-]^{-1}_{\alpha\beta} = \begin{bmatrix} \exp(-i\varphi) & 0 \\ 0 & \exp(i\varphi) \end{bmatrix}, \qquad [R^+]^{-1}_{\alpha\beta} = R^-_{\alpha\beta} = \begin{bmatrix} \exp(i\varphi) & 0 \\ 0 & \exp(-i\varphi) \end{bmatrix}, \quad \text{(S12)}$$

with $\alpha, \beta = \{L, R\}$. In the rotated (primed) coordinate system, the polarization density yields

$$P^{2\omega}_{\alpha'} = \sum_\alpha R^-_{\alpha'\alpha} P^{2\omega}_\alpha = \varepsilon_0 \sum_{\alpha,\beta,\gamma} R^-_{\alpha'\alpha} \chi^{(2)}_{\alpha\beta\gamma} E^\omega_\beta E^\omega_\gamma. \quad \text{(S13)}$$

The field in the primed and unprimed coordinate systems are related as

$$E^\omega_\alpha = \sum_{\alpha'}[R^+]^{-1}_{\alpha\alpha'} E^\omega_{\alpha'}, \quad \text{(S14)}$$

and, substituting (S14) into (S13), we obtain

$$P^{2\omega}_{\alpha'} = \varepsilon_0 \sum_{\beta',\gamma'} \sum_{\alpha,\beta,\gamma} R^-_{\alpha'\alpha} \chi^{(2)}_{\alpha\beta\gamma} [R^+]^{-1}_{\beta\beta'}[R^+]^{-1}_{\gamma\gamma'} E^\omega_{\beta'} E^\omega_{\gamma'}. \quad \text{(S15)}$$

Now we can define the non-linear susceptibility tensor $\chi^{(2)}_{\alpha'\beta'\gamma'}$ in the rotated coordinate system

$$\chi^{(2)}_{\alpha'\beta'\gamma'} = \sum_{\alpha,\beta,\gamma} R^{-}_{\alpha'\alpha} \chi^{(2)}_{\alpha\beta\gamma} [R^{+}]^{-1}_{\beta\beta'} [R^{+}]^{-1}_{\gamma\gamma'}, \tag{S16}$$

and (S15) becomes

$$P^{2\omega}_{\alpha'} = \varepsilon_0 \sum_{\beta'\gamma'} \chi^{(2)}_{\alpha'\beta'\gamma'} E^{\omega}_{\beta'} E^{\omega}_{\gamma'}. \tag{S17}$$

Due to the fact that $R^{\pm}_{\alpha\beta}$ and $[R^{\pm}]^{-1}_{\alpha\beta}$ are diagonal, from (S16) we have

$$\chi^{(2)}_{L'L'L'} = R^{-}_{L'L} \chi^{(2)}_{LLL} [R^{+}]^{-1}_{LL'} [R^{+}]^{-1}_{LL'} = \chi^{(2)}_{LLL} \exp(i3\varphi), \tag{S18a}$$

$$\chi^{(2)}_{L'L'R'} = R^{-}_{L'L} \chi^{(2)}_{LLR} [R^{+}]^{-1}_{LL'} [R^{+}]^{-1}_{RR'} = \chi^{(2)}_{LLR} \exp(i\varphi), \tag{S18b}$$

$$\chi^{(2)}_{L'R'R'} = R^{-}_{L'L} \chi^{(2)}_{LRR} [R^{+}]^{-1}_{RR'} [R^{+}]^{-1}_{RR'} = \chi^{(2)}_{LRR} \exp(-i\varphi), \tag{S18c}$$

$$\chi^{(2)}_{R'R'R'} = R^{-}_{R'R} \chi^{(2)}_{RRR} [R^{+}]^{-1}_{RR'} [R^{+}]^{-1}_{RR'} = \chi^{(2)}_{RRR} \exp(-i3\varphi), \tag{S18d}$$

$$\chi^{(2)}_{R'L'R'} = R^{-}_{R'R} \chi^{(2)}_{RLR} [R^{+}]^{-1}_{LL'} [R^{+}]^{-1}_{RR'} = \chi^{(2)}_{RLR} \exp(-i\varphi), \tag{S18e}$$

$$\chi^{(2)}_{R'L'L'} = R^{-}_{R'R} \chi^{(2)}_{RLL} [R^{+}]^{-1}_{LL'} [R^{+}]^{-1}_{LL'} = \chi^{(2)}_{RLL} \exp(i\varphi). \tag{S18f}$$

The equations (18a)-(18f) can be understood intuitively. A circularly polarized pump wave, when impinging onto a metasurface with spatial variation of element orientation, acquires a spatially varying geometrical phase difference $\pm\varphi(x,y)$ (LHCP or RHCP respectively). Since the nonlinear SH field is proportional to the square of the pump wave, the phase of induced second harmonic field will be double of that, i.e., $\pm 2\varphi(x,y)$. In addition, the second harmonic field acquires an additional phase shift due to scattering of the SH wave on the metasurface with the element orientation gradient, so that the radiated field will have the phase $\pm 2\varphi(x,y) \pm \varphi(x,y)$, depending on the polarization. We are therefore able to locally control the generated nonlinear wavefront with subwavelength resolution and large efficiency. The induced nonlinear polarization current density in the circular polarization basis becomes

$$J^{2\omega}_{\alpha}(\varphi) = -i2\omega P^{2\omega}_{\alpha}(\varphi) = -i2\omega\varepsilon_0 \sum_{\beta\gamma} \chi^{(2)}_{\alpha\beta\gamma}(\varphi) E^{\omega}_{\beta} E^{\omega}_{\gamma}, \tag{S19}$$

where the primes on the indices have been omitted. Explicitly,

$$J^{2\omega}_{L}(\varphi) = -i2\omega\varepsilon_0 \left\{ \chi^{(2)}_{LLL}(\varphi)[E^{\omega}_{L}]^2 + \chi^{(2)}_{LRR}(\varphi)[E^{\omega}_{R}]^2 + 2\chi^{(2)}_{LLR}(\varphi) E^{\omega}_{L} E^{\omega}_{R} \right\}, \tag{S20a}$$

$$J^{2\omega}_{R}(\varphi) = -i2\omega\varepsilon_0 \left\{ \chi^{(2)}_{RLL}(\varphi)[E^{\omega}_{L}]^2 + \chi^{(2)}_{RRR}(\varphi)[E^{\omega}_{R}]^2 + 2\chi^{(2)}_{RLR}(\varphi) E^{\omega}_{L} E^{\omega}_{R} \right\}. \tag{S20b}$$

Since the structure is electrically very thin, we can define an effective surface current $K^{2\omega}_{\alpha}(\varphi)$ as

$$K^{2\omega}_{\alpha}(\varphi) = h J^{2\omega}_{\alpha}(\varphi), \tag{S21}$$

where $h$ is the height of the MQW in the unit cell. Finally, ignoring the phase factor "$-i$" in Eqs. (S20a), (S20b), for a LHCP incident wave we obtain

$$K_{R(L)}^{2\omega}(x, y) = 2\omega\varepsilon_0 h \chi_{RLL}^{(2)\,\text{eff}} [E_L^\omega]^2 \exp[i\varphi(x, y)], \tag{S22a}$$

$$K_{L(L)}^{2\omega}(x, y) = 2\omega\varepsilon_0 h \chi_{LLL}^{(2)\,\text{eff}} [E_L^\omega]^2 \exp[i3\varphi(x, y)], \tag{S22b}$$

and for a RHCP wave we have

$$K_{R(R)}^{2\omega}(x, y) = 2\omega\varepsilon_0 h \chi_{RRR}^{(2)\,\text{eff}} [E_R^\omega]^2 \exp[-i3\varphi(x, y)], \tag{S23a}$$

$$K_{L(R)}^{2\omega}(x, y) = 2\omega\varepsilon_0 h \chi_{LRR}^{(2)\,\text{eff}} [E_R^\omega]^2 \exp[-i\varphi(x, y)]. \tag{S23b}$$

Note that a phase difference between waves scattered from identical optical elements or waveplates at different orientations manifests the fact that the waves traversed different polarization state histories. This effect being of a purely geometrical nature has been first discovered by Pancharatnam and later generalized by sir Berry for quantum mechanical and optical systems undergoing a cyclic sequence of polarization transformations [21,22], and now is often referenced as Pancharatnam-Berry (PB) phase. The classical definition of PB phase is the phase difference between the initial beam and a component of the same polarization in the transmitted field, propagating in the same direction. However, the concept of geometrically induced phase difference can be extended to the case when the same beam passes through waveplates at different orientations which alter the polarization state not in the same fashion. It has been shown that in this scenario the phase difference between these transmitted beams is not zero [23]. In particular, the portions of transmitted waves with the same polarization as the incident wave do not acquire a phase shift, whereas the portions with orthogonal polarization develop a relative phase difference, which is a natural extension of PB phase. Historically, optical elements employing this effect are called Pancharatnam-Berry optical elements [23].

## II. Far Field Calculations for Metasurface with Linear Orientation Variation

In the previous section, we made the assumption that the vertical nonlinear currents may be neglected, i.e., $\chi^{(2)}_{zjk}(\mathbf{r}) = 0$. For non-homogeneous metasurfaces as the ones considered in the paper, however, vertical currents may have an effect, especially when radiation at steep angles from the normal is considered. In order to verify the validity of our assumption and overall accuracy of the presented theory, we compare its results with rigorous numerical simulations for a specific form of transverse inhomogeneity. More specifically, we consider the case, examined also in the main paper, for which the orientation changes linearly with the *x*-coordinate, $\varphi = \varphi(x)$. As outlined in the main paper, this linear rotation is expected to produce circularly polarized beams in specific directions. Since in this example the metasurface forms a periodic superlattice with period $L = Nd$, where $d$ is the size of the unit cell, and $N$ is the number of cells required to achieve a full turn around the $z$-axis, it is actually possible to describe the radiated far-field of the infinite superlattice as a superposition of a finite number of plane waves (diffraction orders) propagating at angles $\theta_n$ (with $\theta_n$ being real), as in a conventional grating. Their complex amplitudes can be calculated again using reciprocity, and they are found proportional to

$$A_{n,\alpha\beta\gamma} = \frac{1}{V}\int_V \chi^{(2)}_{zzz}(\mathbf{r}) f^{2\omega}_{n,\alpha}(\mathbf{r}) f^{\omega}_{\beta}(\mathbf{r}) f^{\omega}_{\gamma}(\mathbf{r}) d^3\mathbf{r}, \tag{S24}$$

in which the indexes correspond to LHCP and RHCP components. The integration is carried out over the volume of the entire *supercell*, and $f^{\omega}_{\alpha}(\mathbf{r})$ and $f^{2\omega}_{n,\alpha}(\mathbf{r})$ are local field enhancement ratios of the $E_z$ component of the field induced in the cell by circular polarized waves $E^{\omega}_{\alpha(\text{inc})}(\theta)$ and $E^{2\omega}_{\alpha(\text{inc})}(\theta_n)$, respectively,

$$f^{\omega}_{\alpha}(\mathbf{r}) = \frac{E^{\omega}_{z(\alpha)}(\mathbf{r},\theta)}{\left|E^{\omega}_{\alpha(\text{inc})}(\theta)\right|}, \qquad f^{2\omega}_{n,\alpha}(\mathbf{r}) = \frac{E^{2\omega}_{z(\alpha)}(\mathbf{r},\theta_n)}{\left|E^{2\omega}_{\alpha(\text{inc})}(\theta_n)\right|}. \tag{S25}$$

Fig. S1 shows numerical results for the amplitude moduli of the propagating diffraction orders evaluated by means of Eq. (S24). The parameters we use are the same as in Fig. 3(d)-(f) of the main paper. One can see that the two largest amplitudes correspond to orders +1 and +3, and they are almost purely RHCP and LHCP polarized, respectively. It should be stressed that a small cross-polarization of the radiated field for the +3 beam is expected even in the ideal case of a purely RHCP current, due to the radiation off the normal.

Other non-zero amplitudes are associated with small, but finite $\chi^{(2)}_{zjk}(\mathbf{r})$ elements, which give raise to the zeroth and $\pm 2$-nd diffraction orders. The rest is due to unwanted variations of the local nonlinear field amplitudes inside each cell, due to the coupling between neighboring elements. Overall, results given by Eq. (S24a), (S24b) agree well with exact numerical simulations, accurately predicting the direction of the two main lobes of the far field. Fig. S2 shows the corresponding exact spatial field distribution for a 30-µm incident RHCP Gaussian beam. The picture is nearly identical to Fig. 3(g) of the main paper, confirming the applicability of our theory. We have carried out similar simulations for several other linear profiles, confirming the presented results.

## III. Radiation from an LHCP Focusing Metasurface Excited with RHCP Waves

The second example in the main paper demonstrates focusing of a LHCP radiated beam at $2\omega$ under LHCP incidence. If the same metasurface is illuminated by a RHCP beam, the phase dependence will be opposite, see Fig. S3 (a), so that instead of focusing the metasurface will show a strong light defocusing, as clearly seen in Fig. S3 (c). In order to achieve focusing under RHCP incidence, the local orientations of PB elements should be reversed.

## IV. Theoretical and Numerical Simulation Methods

Numerical simulations of nonlinear systems can be rather challenging due to instabilities and frequency mixing. The approach implemented here allows the evaluation of the nonlinear response of the system without direct nonlinear simulations, under the undepleted pump approximation. We take advantage of the fact that the second harmonic field is much weaker than the pump and that the nonlinearity is confined in the *zzz* component of the multi-quantum well nonlinear tensor, which allows the use of reciprocity to relate the pump and second harmonic field through the overlap integral over the volume of MQW layers, given by Eq. (S1). A more detailed description of the derivation of Eq. (S1), which is a keystone of our numerical analysis, can be found in [10,29]. To evaluate the local Pacharatnam-Berry phase of a nonlinear gradient metasurface, a unit cell with a particular orientation $\varphi$ is considered to be in a periodic lattice. Then, the structure is illuminated by linearly polarized plane waves impinging normally at frequencies $\omega$ and $2\omega$. This approximation is justifiable here since the difference in orientation between the neighboring elements is small enough and their coupling is weak. These simulations were done in CST Micrwoave Studio [31]. The fields inside the MQW volume of the cell computed at both frequencies were combined according to Eq. (S1) in order to obtain the effective transverse susceptibility components in the Cartesian basis versus the rotation angle $\varphi$. As mentioned above, here we do not account for *z*-oriented nonlinear currents induced in MQW since they do not radiate normally. Then, susceptibility tensor components for elements rotated by a certain angle with respect to a simulated one can be readily found in the circular basis using Eqs. (S10) and (S18). Note that our theoretical model accounts only for the variation of the phase of the induced nonlinear currents, and in practice small variation of the amplitude may occur.

In order to validate our theoretical results numerically, we use the same simulation procedure applied to a finite number of cells (but still arranged in a periodic superlattice) with a linear and quadratic variation of the elements orientation (the two cases considered in the main paper). However, in this case we apply Eq. (S1) to each cell independently, again taking advantage of the fact that the orientation changes slowly from cell to cell, and each one can be thought of as radiating normally. Nevertheless, now each cell feels the difference in the orientation of its neighbors, which affects not only the phase but also the amplitudes, as can be seen in Fig. 3(a),(b),(d),(e), Fig. 4(a),(b), and Fig. S3(a),(b). After this step, the effective transverse nonlinear surface currents excited by a Gaussian beam were computed by means of Eqs. (S22)-(S23), impressed onto a 2D surface, and allowed to radiate using COMSOL [32]. The results of these numerical simulations are shown in Fig. 3(c),(f), 4(c), S2, and S3(c).

## V. SHG Generation Efficiency

One of the goals of the paper is to demonstrate that adding phase control does not require compromising the performance of the SHG process in the proposed multi-quantum-well metasurface. However, since the plasmonic elements provide very large field enhancement, the question of SHG efficiency in such metasurfaces cannot be fully addressed without taking into account saturation effects. As we reported earlier [10,29], the intensity of the second harmonic signal does not grow quadratically with the pump intensity as it would happen in the absence of saturation and losses. In order to explicitly take into account these effects in ultrathin nonlinear metasurfaces we developed a comprehensive theory which closely matches the existing experimental data and allows predicting the maximum possible efficiency of the nonlinear process [29].

For the unit cell considered in this paper (not rotated) the predicted maximal SHG conversion efficiency $\approx 2 \cdot 10^{-4}\%$ is achieved for *y*-polarized pump wave with the intensity of just 10kW/cm$^2$ and *x*-polarized second harmonic wave, see Fig. S4. This result is of the same order of magnitude as the one we reported in [10]. Above this critical intensity, the saturation effects start to dominate and reduce the SHG efficiency. In fact, the actual value of the critical pump intensity depends on multiple factors, the most important among which are the shape of plasmonic resonators and the intrinsic features of the MQW configuration [29]. In the main paper we assume the pump intensity to be well below the MQW saturation intensity [10] and show the magnitudes of effective nonlinear susceptibility tensor components normalized by $\chi^{(2)}_{zzz}$. Though the design of the unit cell is based on a previously reported MQW heterostructure [10] with $\chi^{(2)}_{zzz} = 54\,\text{nm} \cdot \text{V}^{-1}$, the use of higher values will readily increase the conversion efficiency [29]. It is also important to mention that, since we consider circular polarized waves, the SHG efficiency is expected to be somewhat lower than for linearly polarized excitation.

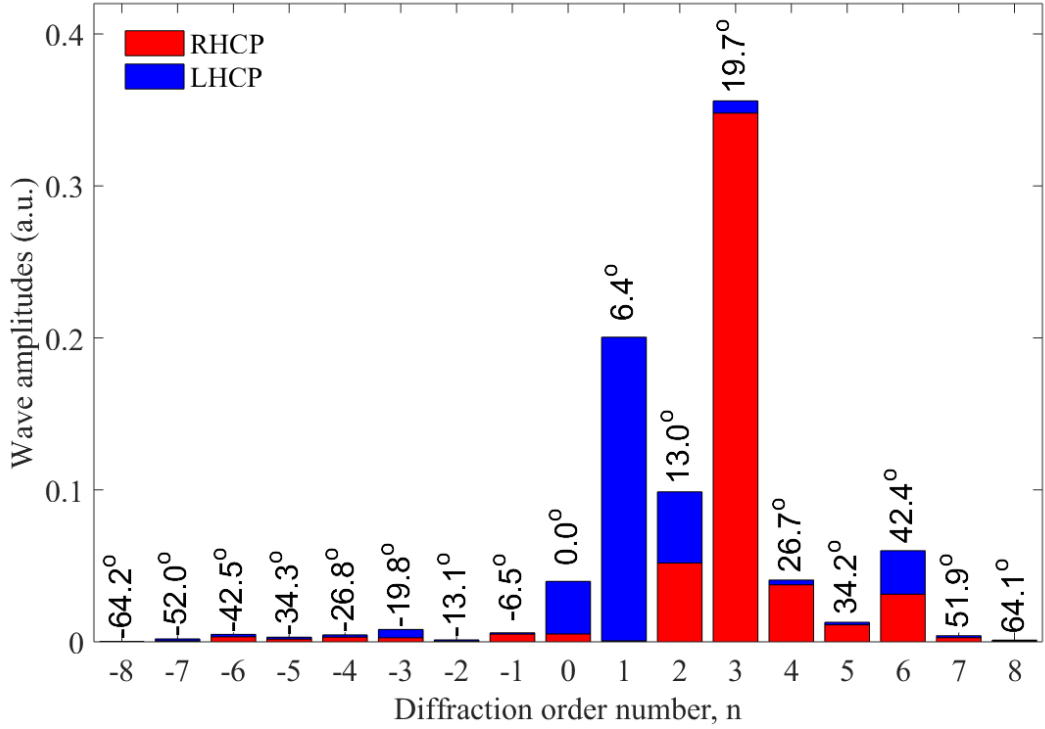

FIG. S1. Numerical results for the amplitudes of propagating diffraction orders for a normally incident RHCP wave. The corresponding propagation angle is indicated above each bar.

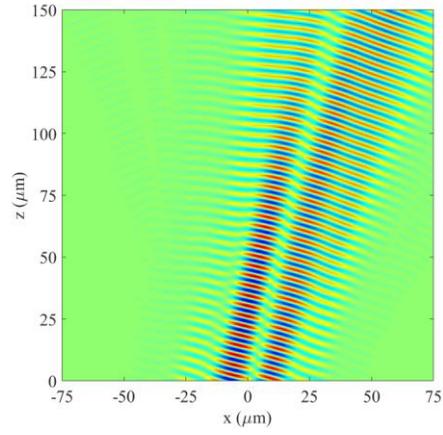

FIG. S2. The field radiated at the second harmonic frequency for a RHCP incident wave with linear rotation (rotation step is 15 deg ).

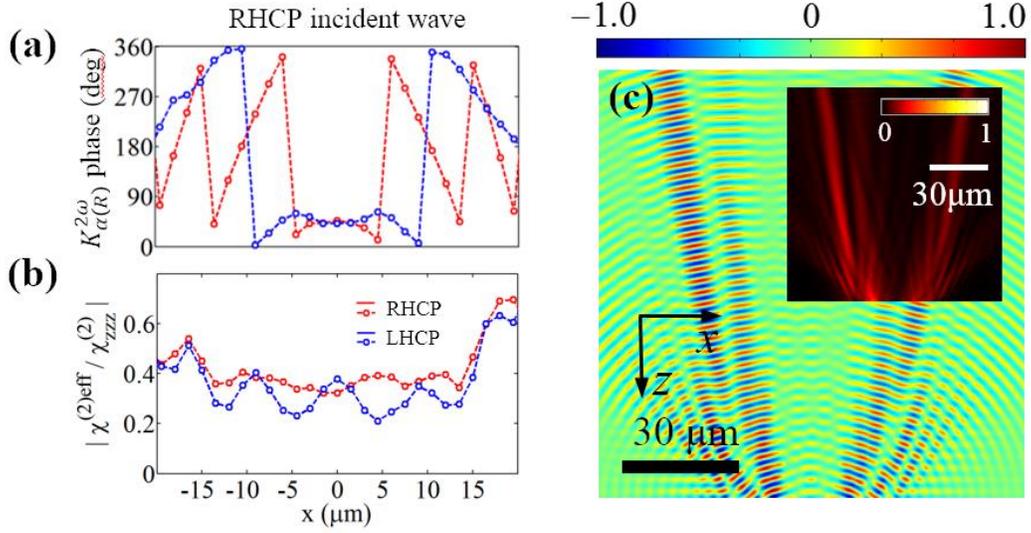

FIG. S3. The structure aimed at focusing the radiated LHCP beam at $20\,\mu m$ from the metasurface under LHCP incidence, but illuminated by a RHCP wave and producing strong defocusing effect. (a) Numerically computed phases of the effective transverse surface currents induced in each cell (see markers) of the metasurface. (b) The magnitudes of the non-zero effective susceptibility tensor components, normalized by $\chi_{zzz}^{(2)}$. (c) The spatial distribution of $E_y^{2\omega}$ above the metasurface. The inset shows the corresponding spatial distribution of the energy density.

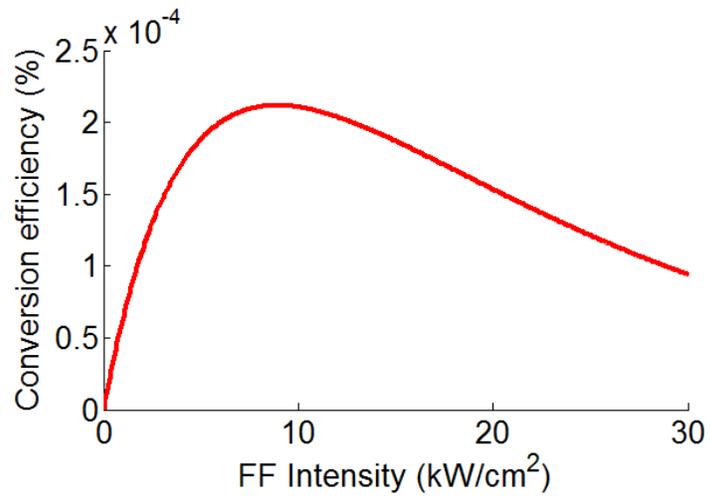

FIG. S4. Efficiency of the SHG process as a function of the pump intensity, taking into account the saturation effects and the presence of losses. The pump wave is linearly polarized along $y$-axis, and the radiated wave is $x$-polarized.